# Moiré Engineering in 2D Heterostructures with Process-Induced Strain

Tara Peña[†,*], Aditya Dey[‡], Shoieb A. Chowdhury[‡], Ahmad Azizimanesh[†], Wenhui Hou[†], Arfan Sewaket[†], Carla Watson[¶], Hesam Askari[‡], and Stephen M. Wu[†,¶,*]

[†]Department of Electrical & Computer Engineering, University of Rochester, Rochester, New York 14627, USA.

[‡]Department of Mechanical Engineering, University of Rochester, Rochester, New York 14627, USA.

[¶]Department of Physics & Astronomy, University of Rochester, Rochester, New York 14627, USA.

**KEYWORDS:**
heterostrain, moiré engineering, process-induced strain, twisted bilayer graphene

**ABSTRACT:**

We report deterministic control over moiré superlattice interference pattern in twisted bilayer graphene by implementing designable device-level heterostrain with process-induced strain engineering, a widely used technique in industrial silicon nanofabrication processes. By depositing stressed thin films onto our twisted bilayer graphene samples, heterostrain magnitude and strain directionality can be controlled by stressor film force (film stress × film thickness) and patterned stressor geometry, respectively. We examine strain and moiré interference with Raman spectroscopy through in-plane and moiré-activated phonon mode shifts. Results support systematic $C_3$ rotational symmetry breaking and tunable periodicity in moiré superlattices under the application of uniaxial or biaxial heterostrain. Experimental results are validated by molecular statics simulations and density functional theory based first principles calculations. This provides a method to not only tune moiré interference without additional twisting, but also allows for a systematic pathway to explore different van der Waals based moiré superlattice symmetries by deterministic design.

When heterostructures of stacked 2D van der Waals (vdW) materials are constructed with a relative twist between the layers, in-plane moiré interference patterns are created that vary in periodicity depending on angle. This in-plane superlattice effect can lead to the formation of flat-bands in twisted bilayer graphene (TBG) near a "magic-angle", which can cause strong electron interactions[1-5]. Within this system alone, an amazing number of strongly correlated electronic phases appear that may be tuned by band filling[1-6]. Within the broader class of twisted vdW systems, including semiconducting transition metal dichalcogenides, there have also been reports of both similar electronic phases and strongly interacting optical phases[7-12].

Critically, these systems rely on the built-in moiré interference patterns formed from only twisting layers. The total space of possible moiré interference patterns is much larger when accounting for strain in each individual layer along with twist (Fig. 1). By individually straining one layer with respect to the other (heterostrain), substantial moiré tunability can be achieved. Specifically, biaxial or uniaxial heterostrain in twisted bilayer 2D heterostructures provides a pathway to manipulate the period of moiré interference (Fig. 1b) or change the geometry of the interference pattern (Fig. 1d). Much of the correlated electronic physics in twisted bilayer vdW systems can be understood in the framework of the Hubbard model, with symmetry of the Hubbard lattice, the Coulomb repulsion (U) and hopping (t) terms all set by twist angle alone[13-15]. A greatly expanded set of available lattice symmetries, and direct control over U/t is possible with deterministic heterostrain in twisted bilayer 2D systems (Fig. 1), allowing an opportunity to better understand correlated electronic phases or serve as a basis for predicting new phases.

One key missing factor in exploring any of these concepts is the deterministic and continuous control of layer-by-layer strain, which has been lacking in the community studying twisted vdW moiré materials[16]. Some works have explored the effects of applied heterostrain through probing uncontrollable heterostrain from fabrication processes[17,18], engineered lattice mismatched epitaxial growth[19], and bending bilayers on flexible substrates[20], but there is a need for a general method compatible with all experimental probes and conditions. Systematic exploration of the strain-expanded phase space of 2D moiré quantum materials requires deterministic application of layer-by-layer strain on the device-level, compatible at low temperatures, with full control over uniaxiality vs. biaxiality. Here, we present a method that meets all the above criteria with the deposition of highly stressed thin film capping layers (process-induced strain). We demonstrate that heterostrain magnitude is directly proportional to the film force (film stress [GPa] × film thickness [nm]) of the stressor layer, and strain directionality is controlled by stressor geometry (Fig. 1a,c).

Process-induced strain engineering has been implemented in industrial silicon technology since the 90 nm technology node, to selectively enhance electron or hole mobility in transistor channels[21-23]. In our previous work, this concept was proven to be equally versatile on vdW bonded 2D materials through evaporating highly stressed thin films onto various 2D multilayer flakes and heterostructures[24-26]. In these works, we show that process-induced strain can allow for the deterministic design over strain magnitude, tension or compression, uniaxiality or biaxiality, and strain direction relative to the crystal axes[24,25]. These methods when applied to 2D materials do not induce damage, are highly time stable and low-temperature compatible[26]. Moreover, we also find incomplete out-of-plane strain transfer (heterostrain) naturally occurs in 2D structures from this method because of their weak vdW coupling[27], allowing for engineered layer-by-layer control over strain in 2D materials.

In this Letter, we apply process-induced strain to controllably heterostrain TBG structures to engineer the moiré periodicity and symmetry. Because bilayer graphene has weak interlayer coupling, we can strain the top graphene layer in this system, while the bottom layer is fixed to the substrate and entirely unstrained[24,28]. This method is general to all 2D materials where substrate adhesion is higher than interlayer adhesion and can be integrated with complex device structures on-chip, meeting all the criteria to examine these systems with low-temperature quantum transport experiments.

Raman spectroscopy is a popular, nondestructive method to extract details about strain, doping, and sample quality in 2D systems such as graphene and $MoS_2$[29,30]. Specifically in high-angle TBG structures, an additional phonon mode is activated by the moiré superlattice, whose phonon frequency can be related back to the twist angle between the two monolayers (Fig. 2a,b)[31]. To understand this mode in TBG, one could look at the superlattice in reciprocal space, where two monolayer graphene Brillouin zones (BZs) are rotated by the given twist angle ($\theta$) with respect to each other. The BZ of the moiré interference pattern (moiré-BZ) has reciprocal lattice vectors $q_1$ and $q_2$ (Fig. 2c), extracted from the difference between the monolayer reciprocal lattice vectors. The TBG's moiré activated phonon mode, termed as the R' (R) band, has a frequency dependent on the magnitude of vector $q_{1,2}$. The R' (R) bands can be understood as the phonon dispersion curves along the Γ-K direction in bilayer graphene, where the longitudinal and transverse optical phonon branches (LO & TO) are active for twist angles below and above ~10° respectively (see supplementary, Fig. 2b)[32,33]. Since the vector $q_{1,2}$ of the moiré-BZ is derived from the real-space superlattice vectors (Fig. 2c), this allows Raman analysis to provide information about strain-transfer and the subsequent real-space changes to the superlattice periodicity simultaneously in these structures. This is illustrated in Fig. 2d, where biaxial heterostrain uniformly increases the magnitude of $q_{1,2}$, while uniaxial heterostrain both increases and decreases the magnitude of $q_1$ and $q_2$ respectively. This leads to the effects seen in R' (R) band manipulation by heterostrain in the following sections.

We initially apply biaxial strain to the top layer in TBG structures (11° < $\theta$ < 16°, 5 TBG samples in total), then probe changes in the in-plane (G-peak) and the superlattice phonon modes (R-peak) with increasing biaxial strain magnitude. Biaxial strain is expected since the thin film stressor would contract to relieve tensile stress from all directions thus transferring biaxial strain to top layer in contact with the stressor, as shown in our previous works[24,25]. Raman spectroscopy is conducted on the TBG/$SiO_2$/Si sample before and after full encapsulation with a highly stressed tensile stressor (Fig. S1). Strain magnitude is directly proportional to thin film force application [N m$^{-1}$] of the stressor layer, we choose to vary this parameter by varying stressor thickness (see supplementary information)[34]. Film force of the stressor layer is quantified via standard wafer curvature methods, then employing the Stoney equation[35,36].

Since biaxial strain is being applied to the top graphene layer, the superlattice is expected to retain $C_3$ rotational symmetry, while the real-space moiré periodicity will decrease, and the magnitude of vector $q_{1,2}$ will increase (Fig. 2d). We quantify phonon shifts as the difference of pre- and post-encapsulation Raman peak positions ($\Delta\omega = \omega_{after} - \omega_{before}$), where we observe a redshift in the R-band and blueshift in the G-band in all samples after thin film stressor encapsulation (Fig. 3a). The redshift in $\Delta\omega_R$ matches the expected increase in $|q_{1,2}|$ and the blueshift in $\Delta\omega_G$ confirms compressive strain transfer to the top graphene layer. Raman spectroscopic maps of the R-band peak positions before and after a 32 N m$^{-1}$ thin film force application on a TBG sample ($\theta \sim 14°$) are shown in Fig. 3b-d, showing good uniformity of heterostrain application. Before encapsulation, the TBG sample displays an average R-band peak position of 1461.1 ± 0.15 cm$^{-1}$. After the 32 N m$^{-1}$ encapsulation, we extract a new average R-band peak position of 1456.7 ± 0.17 cm$^{-1}$. We repeat this procedure on four more TBG samples varying in film force application, where we find the magnitude of both the R-band and G-band shifts scale linearly with thin film force application (Fig. 3e,f), demonstrating direct controllability of heterostrain.

We next calculate the expected phonon frequency shifts with varying biaxial compressive strain magnitudes to the top graphene layer in a simulated 13.2° TBG, using a combination of molecular statics (MS) and density functional theory (DFT) (see supplementary). Slopes of -30.48 cm$^{-1}$/% and 10.21 cm$^{-1}$/% are obtained for $\Delta\omega_R$ and $\Delta\omega_{G,top}$ with applied strain respectively from these calculations. While a peak shift rate of 10.21 cm$^{-1}$/% for the G-band is smaller than literature values for biaxially strained graphene[37], Raman shifts per % strain are unique to both the nature of applied strain and to the 2D material structure itself. Because this is the only time biaxial heterostrain on TBG has been probed with Raman spectroscopy, we utilize our experimental peak shifts to validate and compare with our theoretical calculations. To independently compare experimental peak shifts as a function of strain, we estimate the % biaxial strain in the graphene top layer from film force alone. This is obtained by scaling our own result from biaxially strained monolayer MoS$_2$[24] by the relative elastic moduli between graphene and MoS$_2$. Since graphene has a 3.5 times larger elastic modulus than MoS$_2$, the same film force will generate 3.5 times less strain. The experimental and calculated results are presented together in Fig. 3e,f where there is good agreement between measured peak shifts and simulated values. Thus, we have cross-verified that biaxial heterostrain is being applied to TBG and this directly leads to modification of the moiré-BZ (periodicity) in the expected fashion. We do not expect to resolve the G-peak of each individual layer of graphene, because the G-peaks have a much larger peak width (~12 cm$^{-1}$) than this peak shift itself (~2 cm$^{-1}$), however we do observe this as an increase in the measured G-peak width of ~0.65 cm$^{-1}$.

This good agreement between theory and experiment validates biaxial heterostrain as a mechanism to engineer moiré wavelength, since R-band shifts can directly be mapped back to changes in $\mathbf{q_{1,2}}$, and changes to $\mathbf{q_{1,2}}$ represent changes to real-space moiré interference. These same real space modifications are observed through MS simulations, and subsequently match experimental Raman results one-to-one (Fig. 3e,f). Therefore, these results directly show a closed-loop theoretical/experimental approach to quantitatively engineer moiré periodicity, controlled continuously through biaxial heterostrain.

We now examine designed uniaxial heterostrain to TBG structures. Strain directionality (uniaxiality vs. biaxiality) can be redefined by lithographically patterning the thin film stressors into stripes and modifying the width of the stripe[38]. Uniaxial heterostrain is quite powerful for moiré superlattices particularly because this type of strain can change the geometry of the moiré interference pattern from the typical triangular lattice to more complex patterns (Fig. 1d). To explore this effect, we first lithographically pattern a 2.5 μm wide thin film stressor stripe (40 N m$^{-1}$) onto a 6° TBG sample (Fig. 4b), where the strain is determined to be uniaxial and perpendicular to the direction of the stripe (Fig S5), similar to our results in the past[25].

Prior to thin film stressor deposition, we confirm the TBG sample displays uniform G and R' peak positions throughout the sample (Fig. S3). After a 40 N m$^{-1}$ thin film stripe application, outside the stressor displays the same G and R' positions as before the stressor deposition, then at the center of the stripe we observe both G and R' peak splitting (Fig. 4a). The G peak splitting in this sample arises from heterostrain, where each peak represents the G-peak of an individual layer. This is different than the typical G and G' ($E_{2g}^{+}$ and $E_{2g}^{-}$) peak splitting in individual graphene monolayers due to uniaxial strain that arises from lattice symmetry breaking, since we are well under the limit of strain magnitudes where this is a resolvable effect ($\varepsilon_{uniaxial}$ < 0.6%)[39]. The lower G-peak matches the unstrained (control) G-peak value in the TBG sample of ~1588.5 cm$^{-1}$, therefore represents the signal from the unstrained bottom layer. The upper G-peak is blueshifted by ~10.9 cm$^{-1}$, we can estimate 0.21% compressive uniaxial strain to the top graphene layer when compensating for charge transfer contributions to the G-peak (Fig. S3,4). We can resolve G peak splitting from heterostrain in the uniaxial case (but not the biaxial case), because there is a larger Raman shift rate per % applied strain.

While the G-peak splitting is due to contributions from each individual graphene layer, R'-peaks arise from the combined interaction between both graphene layers. Therefore, the splitting of this peak represents a direct modification to the moiré pattern itself. Under uniaxial heterostrain, we have shown that $q_1$ and $q_2$ vectors both increase and decrease in magnitude, arising from the breaking of the superlattice symmetry of unstrained or biaxially strained moiré interference (Fig. 1d, 2d). Since the R' peak positions are directly related to the reciprocal space lattice vectors for the moiré-BZ ($q_1$,$q_2$), this peak splitting represents direct evidence of $C_3$ rotational symmetry breaking in the moiré superlattice in TBG under deterministic application uniaxial heterostrain. We've seen this repeated across two other samples with patterned stressors and different film forces, where the G and R' peak splittings scale with film force application (Fig. S6).

To quantitatively confirm that the splitting and changes in $\omega_{R'}$ originate from changes in the moiré superlattice symmetry/periodicity, we also conduct Raman spectroscopic mapping over the entire sample (Fig. 4b). When conducting Raman mapping, we observe clear correlations between both emerging R' peaks (Fig. 4c,d) and the $G_{top}$ peak position (Fig. 4e). Strain gradients are found at the edges of the thin film stressor (Fig. 4c-e, S5b), indicating a full range of uniaxial heterostrain engineered by stressor geometry. To examine this more clearly, we plot the R' peak shifts as a function of $G_{top}$ peak shifts (Fig. 4f symbols) and observe a one-to-one linear correlation between the $G_{top}$ and individual R' peak shifts, as expected from a full gradient of moiré interference patterns designed by uniaxial heterostrain within a single sample. We again calculate the phonon responses of a simulated heterostrained 6° TBG structure using MS+DFT to compare with our experimental data, where we find that the calculated phonon shifts match well to the experimental Raman data (dashed lines in Fig. 4f). R' peak shifts per percent strain follow 34.83 cm$^{-1}$/% and -20.51 cm$^{-1}$/%, while the averaged $G_{top}$ peak shift slope is 46.77 cm$^{-1}$/%.

Using our measured $G_{top}$ and R' peak shifts, we can directly back-calculate the **exact** moiré-BZ and visualize the subsequent real-space moiré superlattice at the center of the stressor (Fig. 5b). To illustrate the impact of uniaxial heterostrain application, we present the full evolution of MS+DFT simulated 6° TBG structures in real space under increasing compressive uniaxial heterostrain magnitude from 0% to 1% (Fig. 5). Implementing a combined MS and DFT approach is critical to accurately account for atomic relaxation effects in the simulated heterostrained superlattices[40], allowing us to properly obtain the calculated phonon shifts with applied strain (see supplementary).

Here, we have introduced a method to locally engineer moiré periodicity and symmetry in 2D bilayer heterostructures by design. While this demonstration implemented ~0.2% heterostrain on TBG, this does not represent an upper limit to the amount of strain that can be applied by our method. Several other theoretical and experimental works suggest that ~0.2% heterostrain is already enough to substantially modify the moiré periodicity and subsequently alter the electronic properties in systems like magic-angle TBG[18,19,41-44]. Moreover, we have engineered other 2D systems up to 0.85%[24], with more recent results even reaching 1.5%. The heterostrain applied by process induced strain engineering is general and can be continuously tuned by varying thin film force application and thin film stressor geometry. These heterostrain modifications allow for controlled access to moiré interference patterns that are not achievable through twist alone. Our combined experimental and theoretical approach allows for direct inference of the real space moiré modifications made through heterostrain, allowing for direct quantitative back calculation of the engineered moiré patterns from Raman spectroscopic analysis. This closed-loop approach allows for a systematic engineering of moiré interference in bilayer 2D heterostructures that has not been achieved deterministically before, opening the door to a strain-expanded exploration of twisted 2D quantum materials by design.

**Supporting Information**

Details on the MS simulations, DFT calculations, and experimental Raman methods. Methods of sample preparation, fabrication, and characterization. Quantifying charge transfer from thermally evaporated thin film stressor layer (Fig. S1). Quantifying charge transfer from this multilayer e-beam evaporated stressor (Fig. S2). Raman maps of 6° TBG sample before photolithography and stressor deposition (Fig. S3). Estimating G-peak shift from strain alone with the e-beam multilayer stressor (Fig. S4). Polarization-dependent Raman results and high-resolution line scan over the striped 6° TBG sample (Fig. S5). Raman spectra of a 7° TBG sample before and after patterned thermally evaporated stressor (Fig. S6). Time stability of strain and moiré periodicity on striped 6° TBG sample (Fig. S7).

**Corresponding Authors**

Tara Peña – Department of Electrical & Computer Engineering, University of Rochester, Rochester, NY, 14627, United States; https://orcid.org/0000-0002-8404-9543; email: tpena@ur.rochester.edu

Stephen M. Wu – Department of Electrical & Computer Engineering, University of Rochester, Rochester, NY, 14627, United States; Department of Physics & Astronomy, University of Rochester, Rochester, NY, 14627, United States; https://orcid.org/0000-0001-6079-3354 ; email: stephen.wu@rochester.edu

**ACKNOWLEDGMENT**
We acknowledge support from the National Science Foundation (NSF) (Nos. OMA-1936250 and ECCS-1942815) and the National Science Foundation Graduate Research Fellowship Program (No. DGE-1939268). Raman spectroscopy was performed at the Cornell Center for Materials Research Shared Facilities (CCMR), and CCMR is supported through the NSF MRSEC Program (No. DMR-1719875).

**ABBREVIATIONS**

TBG, twisted bilayer graphene; MS, molecular statics; DFT, density-functional theory; BZ, Brillouin zone; vdW, van der Waals; 2D, two-dimensional.

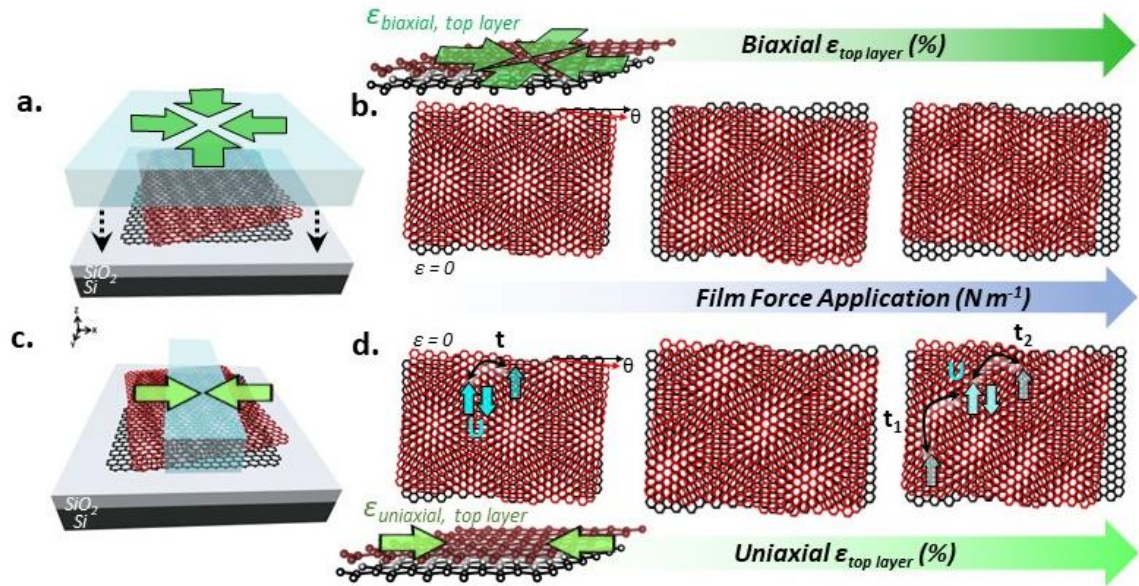

FIG. 1. (a) Highly stressed thin film fully encapsulating a TBG sample, providing biaxial strain to the top graphene layer. (b) Moiré interference pattern varying with biaxial heterostrain magnitude (rigid). (c) Highly stressed thin film patterned into a stripe on TBG, providing uniaxial strain to the top graphene layer (d) Moiré interference pattern varying with uniaxial heterostrain magnitude (rigid) and strain tunable Hubbard model parameters (blue).

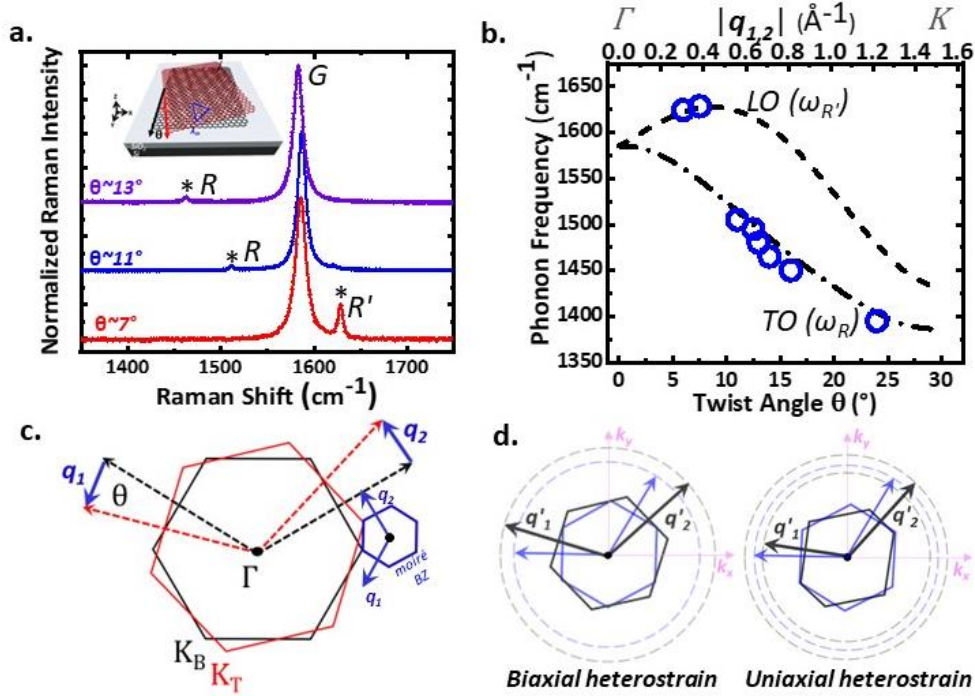

FIG. 2. (a) Raman spectra of TBG samples varying in twist angle. (b) Phonon frequency of the moiré activated phonon mode with twist angle and q-vector magnitude. Open circles represent measured R/R'-bands for various fabricated samples. (c) Brillouin zone schematic of unstrained twisted graphene monolayers (bottom-**black**, top-**red**) and the corresponding moiré Brillouin zone (**blue**). (d) Calculated moiré Brillouin zones under biaxial or uniaxial heterostrain (**grey**) compared to unstrained moiré Brillouin zones (**blue**).

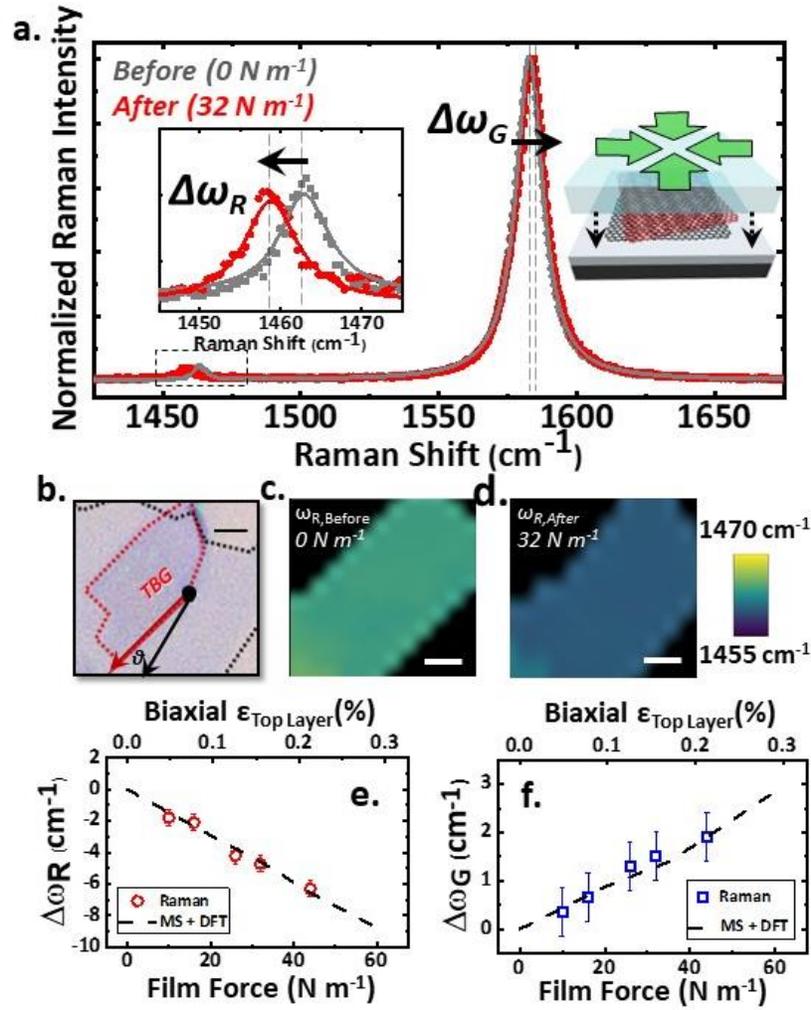

FIG. 3. Biaxial heterostrain case. (a) Raman spectra of a 14° TBG sample before and after a full encapsulation with a 32 N m$^{-1}$ thin film application. (b) Optical micrograph this TBG, scale bar is 1 μm. Raman R-band peak position maps of this sample before (c) and after (d) the full encapsulation, scale bars are 0.5 μm. Film force dependence of $\Delta\omega_R$ (e) and $\Delta\omega_G$ (f).

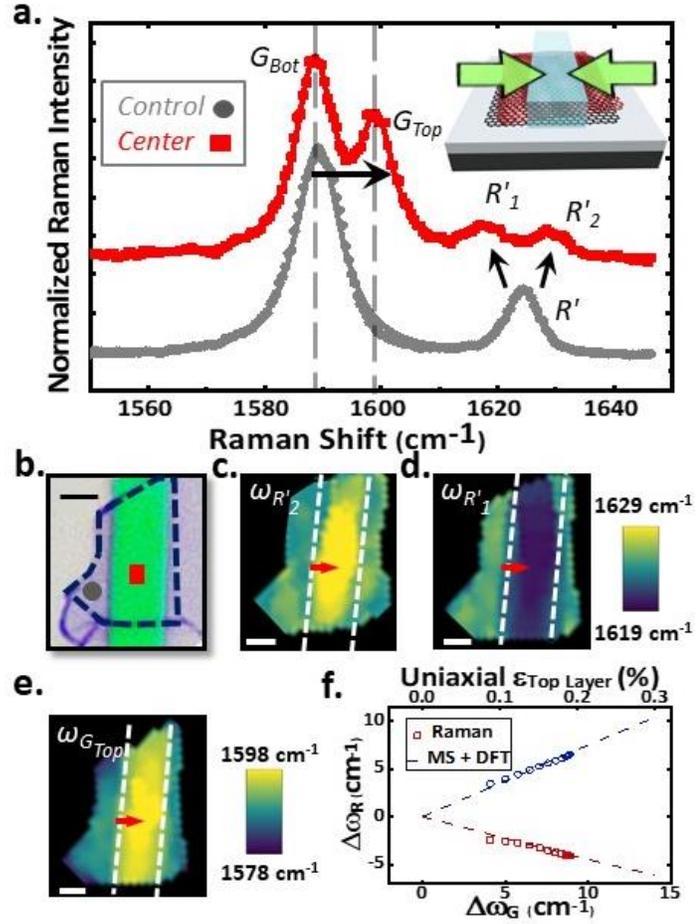

FIG. 4. Uniaxial heterostrain case. (a) Raman spectra outside the stressor and center of the stressor on a striped 6° TBG sample. (b) Optical micrograph of this TBG, scale bar is 2 μm. (c,d) Raman maps of R'$_1$ and R'$_2$ peak positions, with the color bar placed to the right. (e) Raman maps of the G$_{top}$ peak position, with the color bar placed to the right. Raman maps have a 1 μm white scale bar. (f) Δω$_{R'1,R'2}$ as a function of Δω$_{Gtop}$ taken from a line scan (red arrows of Raman maps).

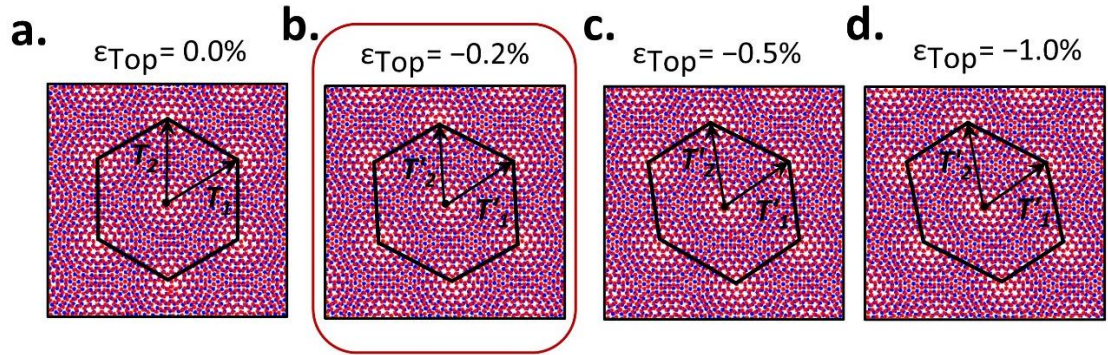

FIG. 5. Evolution of a moiré superlattice with applied uniaxial heterostrain. (a-d) Real-space MS captures of simulated 6° TBG samples varying with uniaxial compressive heterostrain magnitude applied along the zigzag direction.

# Supplementary Information:

# Moiré Engineering in 2D Heterostructures with Process-Induced Strain

Tara Peña[†,*], Aditya Dey[‡], Shoieb A. Chowdhury[‡], Ahmad Azizimanesh[†], Wenhui Hou[†], Arfan Sewaket[†], Carla Watson[¶], Hesam Askari[‡], and Stephen M. Wu[†,¶,*]

*†Department of Electrical & Computer Engineering, University of Rochester, Rochester, New York 14627, USA.*

*‡Department of Mechanical Engineering, University of Rochester, Rochester, New York 14627, USA.*

*¶Department of Physics & Astronomy, University of Rochester, Rochester, New York 14627, USA.*

**Atomistic Simulations**

We performed computational simulations to gain atomistic-level insight of these moiré engineered twisted bilayer graphene (TBG) systems and characterize them to compare it with experiments. Initially, we considered the Bernal stacked (AB stacking) configuration of bilayer graphene and then rotated the top layer by a twist angle θ relative to the bottom layer. This incommensurate stacking orientation of layers results in large periodic supercell of the entire system, i.e., the moiré patterns[1,2]. The size of a moiré pattern resembles the periodic supercell size of a TBG crystal structure. The mathematical equations pertaining to our model is shown in the next subsection. After constructing rigidly twisted lattice structures of TBG (θ = 6° and θ = 13.2°), we performed density functional theory (DFT) simulations to minimize their energy and obtain their relaxed geometries. The minimized lengths of moiré superlattice is 2.3502 nm and 1.0728 nm for 6° and 13.2° TBG systems respectively. Further, to mimic the experimental straining of the system, we conducted molecular statics (MS) simulations (T = 0 K) via energy minimization using LAMMPS[3] based on the optimized supercell structure from DFT (see Superlattice Model). We applied strain (uniaxial/biaxial) on the top layer in a step wise manner (strain is applied along the zigzag direction in the uniaxial case). Since this mechanical deformation is applied to an individual layer, we have taken the film to have a free surface boundary in the loading direction. We observe an oblique patterned distortion of moiré hexagons which further intensifies with strain due to inequivalent magnitude of **q** vectors under uniaxial heterostrain application.

Experimentally obtained Raman peak frequencies can be directly extracted from optical branches of the phonon dispersion spectra of a material along the high symmetry path of its Brillouin zone (BZ)[4,5]. Ab-initio DFT simulations were implemented to compute the phonon spectrum of TBG systems with and without heterostrain. The moiré supercell lattice parameters were calculated for each magnitude of applied strain (see Superlattice Model). By freezing the LAMMPS-simulated atomistic configuration and employing the hetero-strained moiré lattice parameters in the structure, we imported the system for DFT calculations to further minimize the system to first principles-level accuracy, followed by phonon spectrum calculations. As explained by Cocemasov *et al.*, we observe hybrid folded phonon branches in the phonon spectra of TBG because of mixture of different BZ directions of the individual layers[6]. The folded branches were simplified using Phonon Unfolding package[7] as shown in along Γ-K-M-Γ high symmetry path. Raman scattering causes inelastic scattering of photons, due to which atoms vibrate out of phase in the lattice. For analyzing the Raman peaks, we consider the longitudinal (LO) and transverse optical (TO) phonon branches as they depict the out of phase vibrational modes of atoms[8,9].

Unfolded optical branches of TBG resemble that of bilayer graphene and has degenerate phonon bands[6]. With the applied heterostrain, the phonon branches split as a result of electron confinement in the individual layers due to inequivalent strain present in each layer[10-12]. The first-order Raman scattering can be observed from phonons at its BZ center. Hence, the G-band frequency can be obtained from Γ point[5]. At 0% strain, the LO and TO branch converge at a single frequency at the Γ point, which further splits and blueshifts with heterostrain. As the strain is increased, the optical phonon bands of the strained layer further bifurcate corresponding to the doubly degenerate $E_{2g}^+$ and $E_{2g}^-$ phonons[13]. The double resonance Raman mechanism in TBG systems involve scattering by the rotational wavevector **q**, giving rise to R' (R)-band peaks. Using the method demonstrated by Carozo *et al*, we have calculated the R band frequencies for our TBG models from the LO and TO branches[8,9]. The governing mechanism of intra-valley and inter-valley scattering determines the relevant branch for R' (R) peak frequencies (LO θ = 6° and TO for θ = 13.2°). Phonon frequencies corresponding to the magnitude of **q** vectors can be directly utilized to obtain the R band data. The mathematical expressions used to calculate **q** vectors of strained TBG is shown in the Superlattice Model. Like in the G band, we observe splitting of R' band peaks with strain. When uniaxial heterostrain is applied, we obtain unequal magnitude of $q_1$ and $q_2$ vectors. On deriving the R' band peaks corresponding to these vectors for the split optical phonon branches, we observe that with strain the R' band splits into four peaks separated into two branches (two peaks blueshifting and two redshifting). The much smaller splitting within each branch is due to uniaxial strain alone, where symmetry breaking from uniaxial strain in monolayer graphene causes G-peaks to split into $G^+$ and $G^-$ peaks. The same symmetry breaking causes R'-band splitting within each branch, whereas the branches themselves arise from changes in $q_1$ and $q_2$ vectors. Experimentally, at the strain values presented in the main text (< 0.2%), it is not possible to resolve either the G peak splitting from in the top layer from uniaxial strain alone, nor the individual peaks in the two R' branches. The calculated slopes presented in the main text are therefore the averaged blueshifting and redshifting R' peak shifts with applied uniaxial strain, similarly the $G_{top}$ peak shift slope presented is the average of the $G^+$ and $G^-$ ($E_{2g}^+$ and $E_{2g}^-$) peak shifts with applied uniaxial strain (from 0.0% to 0.3%). Simulation and analysis of vibrational properties were conducted up to 0.5% strain, keeping our experimental results in frame. Nonetheless, our computational approach can be used to characterize these systems with a fine control over twist angle and strain.

**Superlattice Model**

The moiré supercell is created by identifying a common periodic lattice for the two layers. Using the mathematical approach demonstrated by Wijik *et al.* and Carozo *et al.*, we have created the real space and BZs for our TBG atomistic model[1,6,8]. To further elucidate the geometrical changes in moiré superlattice, we denote the reciprocal lattice vectors of bottom graphene layer as **b₁** and **b₂** and for the rotated top layer as $\mathbf{b'_1}$ and $\mathbf{b'_2}$. The rotational wavevector or reciprocal lattice vectors of TBG Moire supercell is given as $\mathbf{q}_i = \mathbf{b'}_i - \mathbf{b}_i$ ($i = 1, 2$)[14]. In unstrained configuration, the magnitude of **q** vectors is equal, i. e., |**q₁**| = |**q₂**|. The length of moiré pattern ($\lambda_m$) can be derived using this value of **q** vector as $\lambda_m = \frac{4\pi}{\sqrt{3}\,|\mathbf{q}_i|}$. Now, when strain is applied to the top layer, its reciprocal lattice vector ($\mathbf{b}_i^\varepsilon$) can be mathematically expressed as $\mathbf{b}_i^\varepsilon = (I + S)^{-1}\,\mathbf{b'}_i$, where $I$ is identity matrix and $S$ denotes the strain tensor [15], which can be written as the following for uniaxial compression case, $S = \begin{pmatrix} -\varepsilon & 0 \\ 0 & \nu\varepsilon \end{pmatrix}$. Here, $\varepsilon$ denotes the nominal strain and $\nu$ denotes the Poisson's ratio. The reciprocal lattice vector of TBG with heterostrain can be written as $\mathbf{q}_i^\varepsilon = \mathbf{b}_i^\varepsilon - \mathbf{b}_i$. Like unstrained condition, calculation of $\mathbf{q}_i^\varepsilon$ magnitude shows us that $|\mathbf{q}_1^\varepsilon| \neq |\mathbf{q}_2^\varepsilon|$. Using this set of calculations, we can obtain the size of the moiré supercell with strain and value of **q** vectors to deduce R band frequencies.

**DFT Calculations**

Based on the atomistic model, the real space of our TBG systems were created in ATOMISTIX TOOLKIT (QuantumATK) commercial package[16]. We performed our first principles simulations within the framework of generalized gradient approximation (GGA) embodied in Quantum Espresso open-source package[17,18]. The GGA along with the Perdew-Burke-Ernzerhof (PBE) form has been used as the exchange correlation functional with ultrasoft pseudopotentials[19-22]. The van der Waals interaction is accounted for using the semi-empirical

Grimme functional. The wavefunctions are expanded using plane wave basis set having an energy cutoff and charge density of 35 Ry and 350 Ry respectively. We used 12 × 12 × 1 k-point grid within Monkhorst-Pack[23-25] scheme to sample the TBG BZ. The structures were optimized until all the atomic forces were less than 0.01 eV/ Å. The in-plane lattice constants were relaxed with an out of plane vacuum space of 20 Å to avoid interaction between the periodic images. Phonon dispersion spectra of TBG structures was computed by employing self-consistent density functional perturbation theory (DFPT)[26]. In this method, the dynamical matrices were first computed on a sufficient q-point grid. The inter-atomic constant used in calculating the phonon dispersion and phonon density of states was computed from the Fourier interpolation of the dynamical matrices.

**MS Simulations**

MS simulations conducted done using LAMMPS open-source software[27]. The unstrained, DFT relaxed TBG moiré lattice was transformed to an orthogonal cell with dimensions of 32 nm × 20 nm for both θ = 6° (~ 134 moiré supercells) and θ = 13.2° (~640 moiré supercells) systems respectively. The simulation box is considered with free surface boundary condition in the zigzag direction of TBG structure allowing us to employ strain to one of the layers along that direction. The box is considered periodic along the armchair and out of plane direction. A vacuum space of ~50 Å is inserted along the out-plane-direction to avoid interactions with periodic image. Hydrogen passivation was done along the free surface to obtain the most stable structure. The TBG structures are minimized using a conjugate gradient energy minimization method to have minimum energy configurations. A reactive empirical bond order (REBO) potential was used for the intralayer covalent bonds[28] and for the interlayer van der Waals interaction a registry dependent Kolmogorov-Crespi (KC) potential[29] was selected. As TBG contains different local stacking configurations, an interatomic potential that considers registry different than equilibrium minimum energy stacking is needed[3,30]. Subsequently, we loaded the structure with constant incremental compressive strain to the top layer along zigzag direction. The uniaxial strain was incremented by -0.1% up to final strain of -1%. Between each loading step, the atoms of the top layer were kept stationary at the applied strain level and energy minimization was performed. The snapshots of the structure at different strain magnitudes were taken in Ovito open visualization tool[31].

**Raman Spectroscopy**

Raman spectroscopy is conducted using a WITec Alpha300R Confocal Raman microscope with a 532 nm excitation laser and a diffraction limited spot size (100x objective with 0.9 NA). For the fully encapsulated TBG structures, we conduct Raman spectroscopic mapping with a laser power of 2.5 mW, 1200 l/mm spectrometer grating, and 500 nm step sizes. Since we quantify peak shifts before and after full encapsulation, we use the 521 cm$^{-1}$ Si peak to calibrate our measurements. For the TBG structures with a patterned stressor stripe, we conduct Raman mapping with a laser power of ~4 mW, 1800 l/mm spectrometer grating, and 250 nm step sizes. For the line profile presented in Fig. S5b, we perform a line scan with the same parameters except ~50 nm step sizes are used to resolve the narrow tensile edges more efficiently. For the polarization-dependent Raman data presented in Fig. S5c, the analyzer and incident laser polarization are the same, then the sample is rotated every 20°. The intensity of the $G_{top}$ peak is then normalized to an unstrained reference peak ($G_{bot}$) to compensate for any potential variations in total intensity throughout the measurement. All Raman peak characterizations in this work are done with Lorentzian fittings.

**Sample Fabrication**

All monolayer graphene flakes are created by mechanically exfoliating from the bulk crystal onto thoroughly cleaned 300 nm $SiO_2$/Si substrates. The $SiO_2$/Si substrates are placed into acetone then IPA baths for ~15 minutes each. After the solvent baths, the substrates are placed into a reactive ion etching (RIE) chamber, where a 100 W $O_2$ plasma at 250 mtorr is used to treat the $SiO_2$ surface for 3 minutes. The first set of graphene monolayers are created by directly exfoliating onto the $SiO_2$ immediately after removing the substrates from the RIE chamber (< 1 minute), this is done to enhance the adhesion of the graphene monolayers to the $SiO_2$ surface[32]. The samples that are exfoliated directly after the RIE cleaning are termed the "fixed" graphene monolayers, proper adhesion is also confirmed by subjecting the samples to an IPA ultrasonic bath for 30 minutes. After waiting at least an hour after RIE cleaning, new graphene monolayers can be exfoliated that have less adhesion to the $SiO_2$ substrate (termed as "free" graphene) due to accumulated environmental adsorbates. For all exfoliation steps, the substrates are heated to 100° C for 90 secs while the tape in contact, then the tape is removed after the substrates are cooled.

After monolayers are optically identified along with the dominant exfoliation edge (zigzag) edges, the TBG is then constructed by stacking the "free" graphene on top of the "fixed" graphene via a dry-transfer process. A PC/PDMS dome stamp is used to pick up the "free" graphene first at ~120C, then the "fixed" graphene is placed underneath the stamp with the zigzag edges of both monolayers aligned. After edge alignment, the "fixed" graphene is rotated by the desired twist angle θ, then the "free" graphene layer is placed over the "fixed" graphene at ~100C with a speed maintained below 1 μm/s. The high adhesion at the "fixed" graphene's interface with the underlying $SiO_2$ surface ensures that the "free" graphene is removed from the PC/PDMS stamp, thus leaving a TBG/$SiO_2$/Si structure behind without needing to melt the polymer. Differential interference contrast microscopy and atomic force microscopy are both used to confirm no bubbles are present in the samples. Raman spectroscopy is used to ensure the engineered angle was achieved by probing the R' (R) phonon frequency. High adhesion is the crux for this TBG fabrication; however, this level of adhesion is also necessary to avoid delamination issues after stressor deposition. The TBG fabrication details and adhesion characterization follow the procedures described in our previous work[33].

The biaxially strained TBG samples are fully encapsulated with thermally evaporated stressors of $CrO_x$ (10 nm) / $MgF_2$ (X nm), where tensile film force is adjusted by increasing the thickness of the $MgF_2$ layer alone (40 nm to 125 nm). Tensile film force leads to compressive strain transfer into the top graphene layer. The $CrO_x$ layer is deposited on top of the $MgF_2$ to preserve the stress state in this layer by isolating it from humidity[34]. The depositions begin at $5 \times 10^{-6}$ torr, with deposition rates of 0.5-2 Å/s for $MgF_2$ and ~0.5-0.8 Å/s for $CrO_x$. We ensure that thermal evaporation does not induce any damage by probing monolayer samples for the graphene defect peak and confirm no detectable damage[35]. For the uniaxially strained TBG sample with the patterned stripe, we use a direct-write photolithography lift-off process with S1805 photoresist. The S1805 photoresist is spun at 4000 rpm for 45 seconds then baked at 100°C for 90 seconds before exposure. After exposure, the sample is placed in chlorobenzene for 300 seconds, developed in MF319 for 90 seconds, then rinsed with DI $H_2O$. The stressor used for this sample is e-beam evaporated $Al_2O_3$ (10 nm) / $MgF_2$ (75 nm) / Ti (5 nm). Similarly, the $Al_2O_3$ layer is used to preserve the stress state in the $MgF_2$, while the Ti layer protects the graphene from potential damage of the subsequently deposited layers[36]. We again confirm there are no defects from this process through Raman spectroscopy of the graphene defect peak. E-beam deposition also begins at $5 \times 10^{-6}$ torr, where the deposition rates employed are 0.1-0.2 Å/s for Ti, ~1.5 Å/s for $MgF_2$, and ~0.7 Å/s for $Al_2O_3$.

**Charge Transfer from Thermally Evaporated Stressor**

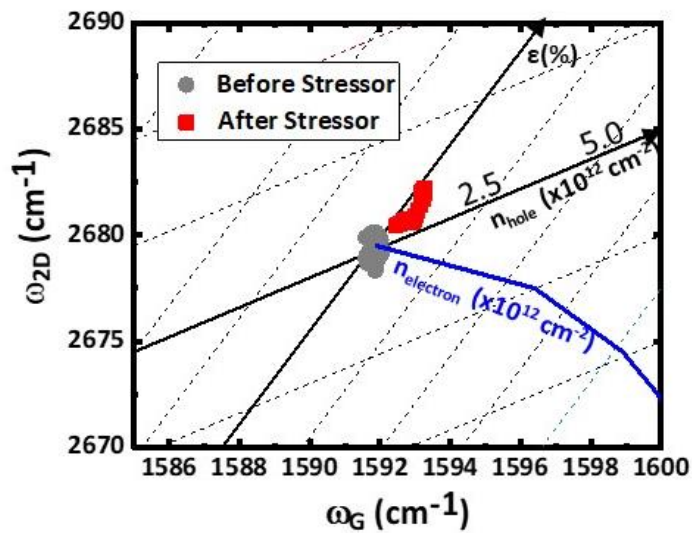

FIG. S1: To quantify the amount of charge transfer from the thermally evaporated $CrO_x/MgF_2$ stressor, we perform Raman mapping before (**grey**) and after (**red**) this encapsulation on a **monolayer** graphene sample. Charge and strain transfer from the stressor can be quantified by knowing $\omega_{2D}$ varies linearly with $\omega_G$ with slopes of 2.2 and 0.7 for strain and charge respectively[37]. An increase in $\omega_{2D}$ and $\omega_G$ is an indication of p-type doping. Here, we find ~$0.1 \times 10^{13}$ cm$^{-1}$ introduced from the stressor encapsulation, also no d-band (no disorder peak) is found from this encapsulation[35]. This is the thin film stressor composition employed in Fig. 3 of the main text, then Fig. S6 in the supplementary.

**Charge Transfer from E-Beam Evaporated Stressor**

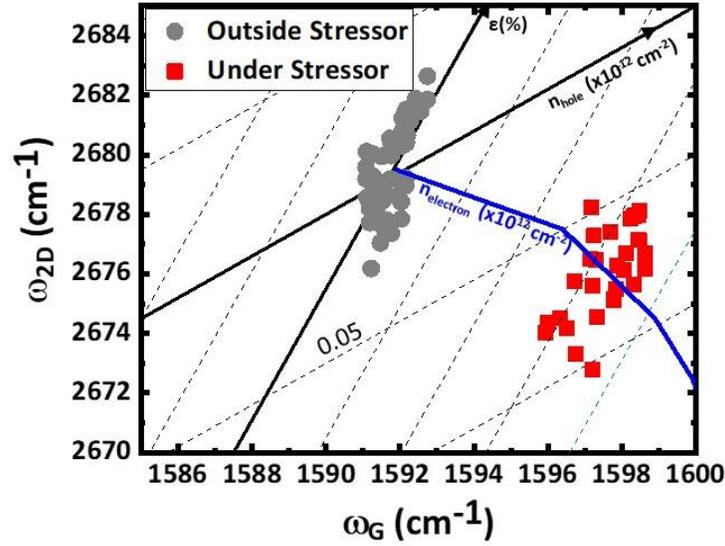

FIG. S2: To quantify the amount of charge transfer from the e-beam evaporated $Al_2O_3/MgF_2/Ti$ stressor, we perform Raman mapping before (**grey**) and after (**red**) this encapsulation on a **monolayer** graphene sample. A decrease in $\omega_{2D}$ and increase $\omega_G$ is indicative of n-type doping[38]. Here, we find $\sim 0.6 \times 10^{13}$ cm$^{-1}$ introduced from the stressor encapsulation and individually confirm an absent disorder peak. We note the thin layer of Ti was used to protect the graphene[36]. This stressor was employed in all data presented in Fig. 4 of the main text.

**Raman Mapping on 6° TBG Without Striped Stressor**

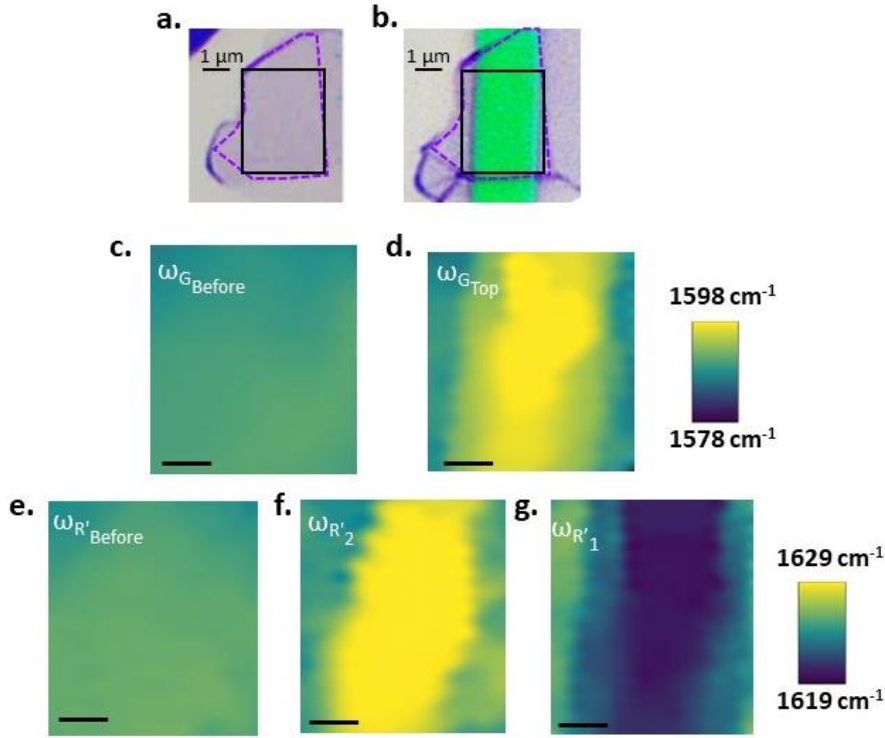

FIG. S3: Optical micrographs of the 6° TBG sample **(a)** before and **(b)** after patterning the striped stressor film. **(c)** Raman map of the G-peak position before the striped stressor. **(d)** Raman map of the $G_{top}$-peak position after the striped stressor. (c) and (d) follow the scale bar to the right. **(e)** Raman map of the R'-peak position before the striped stressor. **(f,g)** Raman maps of the $R'_2$ and $R'_1$ peak positions after the stressor. (e-g) follow the scale bar to the right. All Raman maps have a 1 μm black scale on the bottom left. Here, we confirm none of the effects we observe from uniaxial strain exist on the sample prior.

**Estimating G-Peak Shift from Strain Alone**

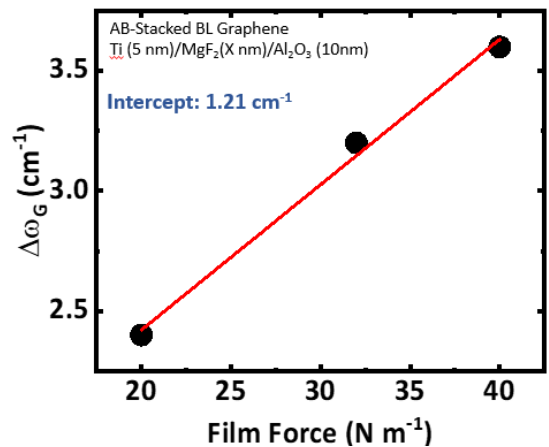

FIG. S4: We extract G-band peak shifts on various AB-stacked bilayer graphene samples with a ~2 μm wide e-beam evaporated stressor, with increasing thin film force. The G-band shifts are extracted from the center of the stressor with respect to the control area. We estimate only an increase of 1.21 cm$^{-1}$ in the G-band comes from charge transfer to the top layer of bilayer graphene samples.

**Polarization-Dependent Raman on the Striped 6° TBG Sample**

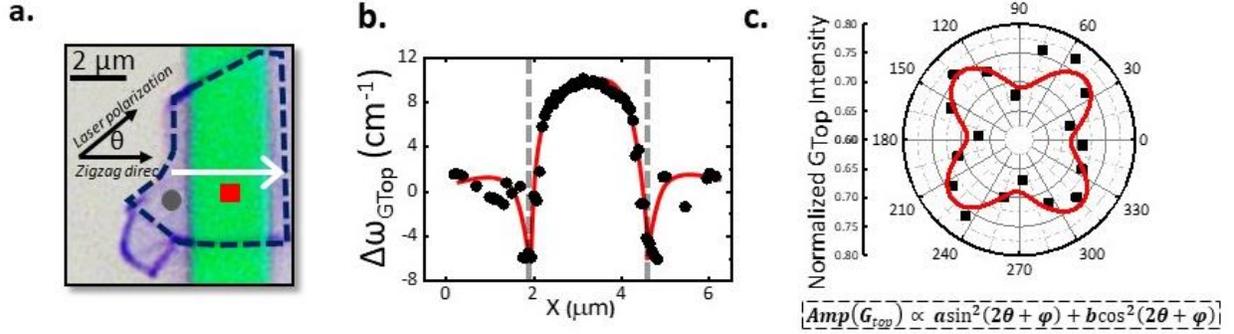

FIG. S5: **(a)** Optical micrograph of the same sample from Fig. 4 of the main text. **(b)** High-resolution line scan through this sample (following white line in (a)). **(c)** Polarization-dependence of the normalized $G_{top}$ peak presented in Fig. 4a. The red solid line is the fit to the corresponding equation, where we find $a \neq b$ and therefore indicates uniaxial strain (or crystal symmetry breaking)[39,40].

**Raman Spectra on TBG Sample Before & After Patterned Thermally Evaporated Stressor**

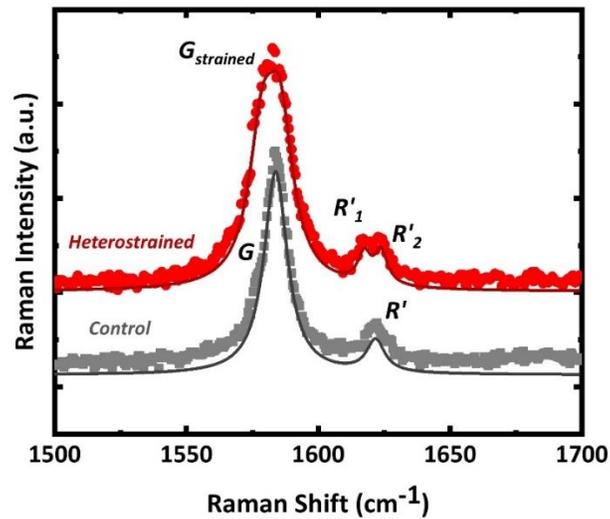

FIG. S6: Raman spectrum of a TBG ($\theta\sim7°$) before (**grey**) and after (**red**) thermally evaporated stressor of $CrO_x$/ $MgF_2$ (18 N m$^{-1}$) patterned onto the sample. The G-peak after the stressor shows peak broadening from heterostrain application (top strained layer and bottom unstrained layer), moreover there is R' peak splitting clearly observed from anisotropic heterostrain.

**Time Stability on Striped 6° TBG Sample**

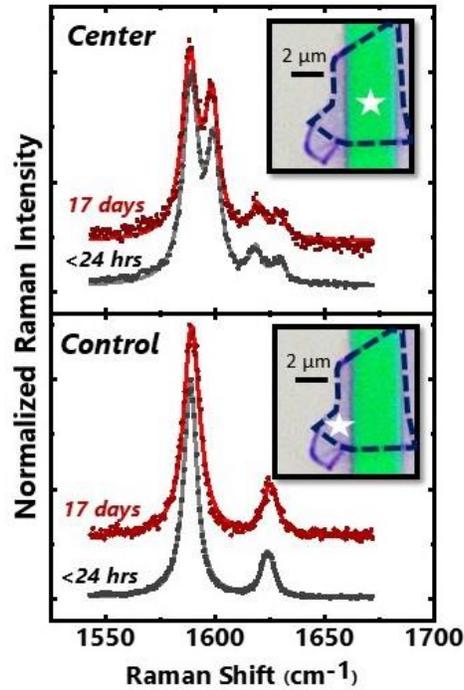

FIG. S7: Raman spectra of the TBG sample from Fig. 4 of the main text with respect to time after stressor deposition. Upon comparing the Raman spectra of <24 hours and 17 days after stressor deposition at the **center of the stressor stripe**, we find both pairs of G and R' peak positions present negligible differences. This suggests that the engineered moiré periodicity is maintained since the strain state itself is. Similarly, at the control regions, there are little differences between the Raman modes with respect to time.